\documentclass[review]{elsarticle}

\usepackage{xcolor}
\usepackage{hyperref}

\journal{Neurocomputing}

\usepackage{numcompress}

\begin{document}

\begin{frontmatter}

\title{Persistence of hierarchical network organization and emergent topologies in models of functional connectivity}

\author[FAUaddress]{Ali Safari}

\author[FAUaddress]{Paolo Moretti\corref{mycorrespondingauthor}}

\cortext[mycorrespondingauthor]{Corresponding author}
\ead{paolo.moretti@fau.de}

\author[ib1,ib2]{Ibai Diez}

\author[cruces,medicine,ikerbasque]{Jesus M. Cortes}

\author[EFMaddress,IC1address]{Miguel \'Angel Mu\~noz}

\address[FAUaddress]{Institute of Materials Simulation, Friedrich-Alexander-Universit\"at Erlangen-N\"urnberg, Dr.-Mack-Str. 77, D-90762 F\"rth, Germany}
\address[ib1]{Functional Neurology Research Group, Department of Neurology, Massachusetts General Hospital, Harvard Medical School, Boston, MA 02115, USA}
\address[ib2]{Gordon Center, Department of Nuclear Medicine, Massachusetts General Hospital, Harvard Medical School, Boston, MA 02115, USA}
\address[cruces]{Biocruces-Bizkaia Health Research Institute, Barakaldo, Spain}
\address[medicine]{Departament of Cell Biology and Histology.  University of the Basque Country (UPV/EHU), Leioa, Spain}
\address[ikerbasque]{IKERBASQUE: The Basque Foundation for Science, Bilbao, Spain}
\address[EFMaddress]{Departamento de Electromagnetismo y Física de la Materia, Universidad de Granada, Granada E-18071, Spain}
\address[IC1address]{Instituto Carlos I de Física Teórica y Computacional, Universidad de Granada, Granada E-18071, Spain}

\begin{abstract}
Functional networks provide a topological description of activity patterns in the brain, as they stem from the propagation of neural activity on the underlying anatomical or structural network of synaptic connections. This latter is well known to be organized  in hierarchical and modular way. While it is assumed that structural networks shape their functional counterparts, it is also hypothesized that alterations of brain dynamics come with transformations of functional connectivity. In this computational study, we introduce a novel methodology to monitor the persistence and breakdown of hierarchical order in functional networks, generated from computational models of activity spreading on both synthetic and real structural connectomes. We show that hierarchical connectivity appears in functional networks in a persistent way if the dynamics is set to be in the quasi-critical regime associated with optimal processing capabilities and normal brain function, while it breaks down in other (supercritical) dynamical regimes, often associated with pathological conditions. Our results offer important clues for the study of optimal neurocomputing architectures and processes, which are capable of controlling patterns of activity and information flow. We conclude that functional connectivity patterns achieve optimal balance between local specialized processing (i.e. segregation) and global integration by inheriting the hierarchical organization of the underlying structural architecture.

\end{abstract}

\begin{keyword}

brain networks \sep functional connectivity \sep hierarchical modular networks \sep segregation integration

\end{keyword}

\end{frontmatter}

\section{Introduction}

The recent convergence of neuroscience and network science opens up snew opportunities to approach the study of brain function \cite{Sporns2004,Bullmore2009,sporns2010}. A fundamental issue in this context is how \textit{structure} and \textit{function} are related \cite{Zhou2006,MullerLinow2008,Bullmore2012,Sporns2013,Diez2015} . In the context of network neuroscience, structure refers to the network mappings of the brain, also known as connectomes, as derived from the actual anatomical connections between  brain regions, also called ``connectomes''  \cite{craddockMACRO,fornitoBOOK}.  Structural connectivity (SC) is thus encoded in networks where nodes are coarse-grained representations of specific brain regions, and the links express the presence of white-matter based connections between pairs of nodes, while weights associated with links conventionally measure the number of such connections. Structural networks are then represented by weighted, non-negative and sparse adjacency matrices, effectively containing the anatomical routing information of a brain. The direction of each connection can currently be recorded only for certain types of connectomes (e.g. mice or primates); in most cases, including that of the human connectome, adjacency matrices are symmetric as directions of connections are still not detected.

Functional connectivity (FC), instead, is often measured from neural activity correlations rather than actual anatomical pathways. As activity correlations can be measured between any pair of nodes, the matrix representations for functional networks differ substantially from those for structural networks in that they lose the property of being sparse. FC matrices are dense and corresponding sparse-network representations can only be obtained by applying arbitrary thresholding procedure. As a further source of complication, FC matrices lose the non-negativity of SC matrices too, as activity correlations are signed, revealing the existence of both correlations and anti-correlations
\cite{anticorrelations}. This aspect makes the application of thresholds an even more delicate issue, subject to a large deal of arbitrariness, as in choosing for instance to exclude anti-correlations, or to study correlations and anti-correlations separately. 

We note that in spite of this complexity, FC data have been recorded for years now and have led to ground-breaking advances in the the understanding of brain function. At the clinical level, FC data have been successfully related to the occurrence of brain pathologies \cite{Biswal2010,Fox2010,VanEssen2012,Lee2013fMRI}. At the graph theoretical level, FC networks have facilitated the introduction of statistical physics concepts such as scale invariance, avalanches, criticality and localization in the brain \cite{Haimovici2013,Friedman2013,Moretti2013,Odor2014,Odor2015,Safari2017}. Advances in acquisition and analysis of FC data have allowed the introduction of novel and challenging concepts, such as that of dynamical FC, i.e., the more ambitious  study of the time dependence and dynamics of functional networks, obtained through the recording of multiple functional networks, each one over a short time window \cite{Allen2014,cabral+Deco,dynamic}. In all these approaches, the question  of what relevant  topological information is lost when thresholds are applied remains  open.

Structural brain networks \cite{Sporns2005,Hagmann2008} are well known examples of biological systems, which exhibit a hierarchical modular structure. Hierarchical modular organization is often described by simple mathematical models of synthetic hierarchical modular networks (HMNs). Is it possible to say the same about functional networks? In principle, the answer to this question is affirmative as also FC networks are known to exhibit modular
and hierarchical patterns of activity \cite{Kaiser2011,fornitoBOOK,cabral+Deco}.
But, how general can the answer to the previous question be if functional network topology can vary so wildly because of arbitrary threshold choices? How persistent is the hierarchical organization of FC data, upon applying different threshold and, more importantly, in different states of neural activity? Are structural differences between SC and their corresponding FC networks representative of pathological states? Answering these questions  is a necessary step towards the understanding of the relationship between structural and functional connectivity in the brain. In particular, the study of this relationship often relies on assessing the similarity between structural and functional connectivity. Similarity measures were proposed in the past (see for instance \cite{Diez2015}), highlighting how ``similarity'' itself may be non-trivially dependent on the choice of thresholds used to extract functional networks. In this work, we follow a more fundamental approach, in the attempt to formulate a criterion to decide if a functional network is hierarchical and to do so irrespective of threshold choices. 

The interest in the hierarchical organization of brain activity patterns is rooted in the observation that hierarchical (and hierarchical-modular) networks exhibit desirable properties of robustness. Remarkably, hierarchical networks do not exhibit a single percolation threshold, where a giant component emerges together with a scale-free distribution of cluster sizes \cite{Dorogovtsev2008}. Instead, in hierarchical modular networks, upon varying the control parameter $p$ (i.e. upon removing a fraction $1-p$ of links or nodes) power-law distributions of connected component sizes are encountered for a broad interval of values of $p$  \cite{Boettcher2009,Friedman2013}. The functional counterparts to this simple structural property are striking:

\begin{itemize}

\item  activity in hierarchical models of the brain is sustained even without any fine-tuning of regulatory mechanisms \cite{Kaiser2007}; 

\item  avalanches of neural activity are power-law distributed in size without the need to fine tune spreading control parameters \cite{Friedman2013,Moretti2013}, rationalizing the experimental observation of scale invariant activity patterns \cite{Beggs2003};

 \item simple dynamical models, normally displaying clear-cut tipping points (phase transitions), exhibit instead \emph{extended quasi-critical phase} (e.g. Griffiths phases \cite{Vojta2006,Munoz2010,Juhasz2012,Moretti2013,Villa2015,Odor2014,Odor2015}), where activity propagates by rare region effects and states of local coherence emerge as chimera-like states \cite{Villegas2014,Millan2018,Millan2019}.
\end{itemize}

 These observations corroborate the view that the hierarchical organization of the structural network of anatomical connections  is responsible for the brain ability to localize activity, avoiding the opposing tendencies where active states die out (as encountered in advance stages of some neurodegenerative diseases)   or invade the system (as typically occurs  during pathological epileptic seizures). From the perspective of neurocomputing and information processing, HMNs thus provide an optimal architecture, which ensures the balance of activity segregation and integration \cite{Gallos2012}, and results in enhanced computational capabilities, large network stability, maximal variety of memory repertoires and maximal dynamic range \cite{Moretti2013}.

  In order to analyze the topology of the resulting FC networks, we exploit their associated spectral properties, allowing us to properly identify {\it hierarchical functional networks}. In particular, we focus on the well-known property of vanishing spectral gaps ---which, following the ideas in \cite{Moretti2013}--- we use as a measure of the hierarchical organization.
Networks characterized by dense connectivity and high synchronizability exhibit a large separation between the two largest eigenvalues of the adjacency matrix \cite{Farkas2001}, or the two  lowest distinct eigenvalues of the Laplacian matrix, a property that we refer to as large spectral gaps. Hierarchical networks of interest in brain modeling, instead, display localized patterns of activity and, as a spectral counterpart, vanishing (although non-zero) spectral gaps \cite{Moretti2013,Villegas2014}. While the spectral characterization of HMNs includes many complex aspects, the single fact that they exhibit vanishing spectral gaps is quite remarkable and has been directly related with some important dynamical features such as self-sustained activity and local coherence \cite{Moretti2013,Villegas2014}. These aspects make hierarchical networks unique and, more importantly, constitute the reason why hierarchical organization is essential in brain networks.

 In what follows, we generate FC data from Monte Carlo simulations of simple models of activity spreading in synthetic HMN models for structural networks and on real SC data. We show that, independently of the choice of thresholds, functional networks generated in the quasi-critical regime always display hierarchical organization (as quantified through spectral gaps). This feature is lost as soon as FC data is produced in the super-critical regime -- i.e. {\it above} quasi-critical regime. Our results offer important insights regarding the topological changes that FC data undergoes in pathological states (the super-critical regime), providing clues for FC analysis as a diagnostic tool. At the same time, our methodology provides a new tool for the analysis of persistent features in functional data, whose appearance is not an artefact of an arbitrary threshold choice.

\section{Materials and methods}\label{sec:mm}

\paragraph{Structural networks} We use computer generated models of HMNs, in order to mimic realistic structural networks.  As different types of algorithms are used in the literature to generate HMNs \cite{Kaiser2007,Wang2011,Moretti2013,Odor2015,Safari2017}, in the following we refer the model proposed in \cite{Safari2017} without loss of generality. We call $\alpha$ the {\it connectivity strength} of a HMN, as it is the main parameter controlling the emerging topology.  The network is organized in densely connected modules of size $M_0$, which represent the level $0$ of a hierarchy of links. At each hierarchical level $i>0$, super-modules of size $M_i=2^iM_0$ are formed, each one joining two sub-modules of size $M_{i-1}$ by wiring their respective nodes with probability $\pi_i=\alpha /4^{i}$: the average number of links between two modules at level $i$ is $n_i=\pi_iM_{i-1}^2=\alpha (M_0/2)^2$, i.e. proportional to $\alpha$, regardless of the value of $i$. For a generic dynamic process running on a HMN in the $N\to \infty$ limit, the effect of lowest-level modules becomes negligible (relegated to transient time scales) and the time asymptotics are dominated by the hierarchical organization: in this regime $\alpha$ becomes the only relevant construction parameter. We remark that, being $n_i$ the number of links between any two modules of size $M_{i-1}$, the highest possible $n_i$ is given by $M_0^2$, so that $\alpha$ can take values in the interval $4/M_0^2\leq \alpha \leq 4$.

\paragraph{Graph spectra} We identify an unweighted graph through its adjacency matrix $A_{ij}$, whose generic element $ij$ is $1$ if nodes $i$ and $j$ are connected, and $0$ otherwise. This choice represents a simplification of the more complex case of a weighted graph, where each link between $i$ and $j$ comes with a weight $W_{ij}$.  The spectrum of such an unweighted graph is given by the eigenvalues of $A_{ij}$. Since we deal with graphs that come in a unique connected component, the Perron-Frobenius theorem ensures that the eigenvalue of largest modulus $\lambda_1$ is unique, real and positive. The spectral gap $g$ is then defined as the difference in modulus between $\lambda_1$ and the second eigenvalue $\lambda_2$. Here we deal with undirected HMNs, so that $A_{ij}$ is symmetric, and $g=\lambda_1-\lambda_2$. We can also compute the Laplacian spectrum of a graph, which consists of the eigenvalues of the graph Laplacian $L_{ij}=\delta_{ij}\sum_l k_j - A_{ij}$ ---where $k_i=\sum_j A_{ij}$ is the degree of node $i$---
 which provides a discretization of the Laplace-Beltrami operator  \cite{FanChung}. If $A_{ij}$ is symmetric, $L_{ij}$ is symmetric and positive semi-definite, and the spectral gap $g_L$ is defined as the modulus of its smallest nonzero eigenvalue. An alternative discretization of the Laplace-Beltrami operator is provided by the normalized Laplacian matrix $\mathcal{L}_{ij}=\delta_{ij}-A_{ij}/\sqrt{k_ik_j}$, for which the spectral gap $g_\mathcal{L}$ can be computed as above.  Finally, for completeness, let us also mention that the results for $\mathcal{L}$ also extend to the random walk Laplacian, which enters the master equation of random walks,  and has the same spectrum of eigenvalues as $\mathcal{L}$ \cite{FanChung}.

\paragraph{Dynamical model}  We simulate dynamics in HMN models of the brain using a null model of activity propagation in network 
---whose use 
has proven effective as a probing tool to understand paramount features of brain activity \cite{Kaiser2007,Moretti2013};--- in particular, we consider the, so-called, susceptible-infected-susceptible (SIS) 
  dynamics. In this well-known model, nodes can be either active (infected I) or inactive (susceptible S); in terms of neural dynamics an active node corresponds to an active region in the brain, which can activate an inactive neighboring region with a given probability $\kappa$, and which can be deactivated at rate $\mu$ (which we set to $1$ without loss of generality) due to e.g., exhaustion of synaptic resources. 

In particular, within the so-called quenched mean-field approximation, the equations for our model read \cite{PastorSatorras2015}
\begin{equation}
\frac{d}{dt}\rho_i(t) = -\mu \rho_i + \kappa [1-\rho_i(t)]\sum_{j=1}^{N} A_{ij}\rho_j(t),
\end{equation}
where $\rho_i(t)$ is the probability that node $i$ is in the active state at time $t$.  We call $\rho(t) = \langle \rho_i(t) \rangle$ the average activity at time $t$. In the general case, SIS dynamics results in a critical value $\kappa=\kappa_c$, above which activity invades the system indefinitely reaching a non-zero steady state $\rho(t\to\infty)>0$ (super-critical regime, in the following). In the case of HMNs, values  $\kappa < \kappa_c$ (quasi-critical regime) constitute a Griffiths phase and are associated with rare-region effects and e.g. slow activity relaxation, pointing to an effective model for normal brain function. We consider initial states where all nodes are active  and simulate time evolution for different values of $\kappa$. Following this protocol in HMNs, simulations are characterized by an initial transient regime, after which, the asymptotic behavior takes over, namely a steady state for $\kappa > \kappa_c$ and a power-law time-decay within a finite range of $\kappa$ below $ \kappa_c$ (a Griffiths phase; see \cite{Munoz2010,Juhasz2012,Moretti2013}).

\paragraph{Generation of functional data} As a proxy for empirically measured FC networks, in our computational study we determine co-activation matrices  \cite{Huett2014,Damicelli2019}. In particular, we acquired activity data from a time interval $I$ in our simulation time series $\rho_i(t)$. In particular, we choose $I$ to occur after the initial transient regime described above (which can be simply determined by visual inspection). If $\kappa > \kappa_c$, $I$ is an interval of the steady state. If $\kappa < \kappa_c$, $I$ covers a part of the power-law-decay  regime instead. From the data in $I$, we generated a matrix $C_{ij}$ with generic $ij$ element equal to the probability that nodes $i$ and $j$ are simultaneously active in a given time bin $\Delta t$ ($\rho_i(\Delta t)=\rho_j(\Delta t)=1$).
Using these data, we averaged $C_{ij}$ across multiple realizations of the dynamics (at least $10^5$ in all cases), and considered a threshold $\theta$, thus generating an adjacency matrix of the functional network for each $\theta$ by imposing $A_{ij}(\theta)= \mathrm{H}(\theta-C_{ij})$, where $\mathrm{H}(\;)$ denotes the Heaviside step function.
Let us remark that, when recording these synthetic functional data, we focused on a single realization of the structural network hosting the dynamics, mimicking a real-life scenario in which diagnostics are conducted on a single subject. 

We us emphasize that the use of co-activation matrices as a proxy for FC is motivated by the fact that in the simplified dynamic model time series are sequences of binary states ($1$ and $0$), which makes it natural to define simultaneous activity. Experimental time series, instead, are not constituted by binary states and require functional data to be computed as Pearson correlation coefficients. As both methods consist in computing activity correlations, we believe that our forthcoming conclusions for the synthetic case and its co-activation matrices carry over to real systems and correlation matrices. Furthermore, since the method proposed here requires only a finite time interval $I$, it finds a natural application to the study of dynamical FC, where functional networks are generated from short time intervals.

\paragraph{Mapping to percolation problem}
For every functional network characterized by an adjacency matrix $A_{ij}(\theta)$, we varied $\theta$ continuously and we monitored how the network structure evolves as measured by two indicators:
\begin{itemize}
 \item the size of the largest connected component $s_1$; 

\item the spectral gap $g$ (as well as its Laplacian counterparts for completeness). 

\end{itemize}

For $\theta\approx 0$ the networks is fully connected, i.e. $s_1=N$ and $g=N$ ($g_L=N$, $g_\mathcal{L}=1$),
 while for $\theta\approx 1$ the network is completely fragmented, i.e., $s_1=1$ and $g\approx 0$ ($g_L=g_\mathcal{L}=0$). Thus, for intermediate $\theta$ the network necessarily undergoes a percolation-like phase transition, signalled by a decrease in $s_1$. If the functional network is non-hierarchical, at the same transition point there is a change in terms of $g$, i.e. its becomes very  small (see below). In other words, both $s_1$ and $g$ act as  order parameters of the same percolation-like transition and share the same {\it critical} value of $\theta$. However, this scenario changes for hierarchical networks. Since a hierarchical network must possess a small spectral gap without being fragmented, the two transitions must be well separated instead. In particular, we expect a threshold interval in which the spectral gap $g$ decreases by orders of magnitude (the order of $N$), while $s_1$ remains close to its maximum. Let us remark that the idea of analyzing the dependence of $s_1$ on $\theta$ in functional networks is not new, but it was developed by Gallos and collaborators \citep{Gallos2012}. However, here the focus is instead on a similar procedure for both $g$ and $s_1$ and comparting the corresponding transition points.

\paragraph{Experimental data}

Even if the present work relies on the generation of functional networks by simulating activity on synthetic structural networks, in order to achieve more general conclusions, we also extended the analyses to the case in which the underlying structural networks are actual connectomes as derived from neuroimaging studies. In particular, we use the connectomes of two healthy subjects, taken from the results of a broad experimental study,
which is described in the Appendix.

\section{Results}

Our main goal is to find an effective way to assess if a functional network is hierarchical, thus mimicking the topology of the underlying structural network, or if activity correlations result in a different, emergent topology. To this end, we define the concept of hierarchical functional network as follows. Since many of the desirable dynamic properties of hierarchical modular networks (rare region effects, generic criticality, localization \cite{Friedman2013,Moretti2013,Villegas2014,Odor2015}) can be traced back to their small spectral gaps $g$ (as well as $g_L$ and $g_\mathcal{L}$), we define a hierarchical functional network as a network that 

\begin{itemize}
\item i) is generated by activity patterns on a structural HMN; 

\item ii) possesses vanishing spectral gaps for a finite range of $\theta$ values, while still being connected (i.e. not fragmented).  
\end{itemize}

Note that the second requirement ensures that the vanishing spectral gap property is not a trivial consequence of a network breaking into many connected components \cite{FanChung}.

\begin{figure}
	\centering
	\includegraphics[width=.95\textwidth]{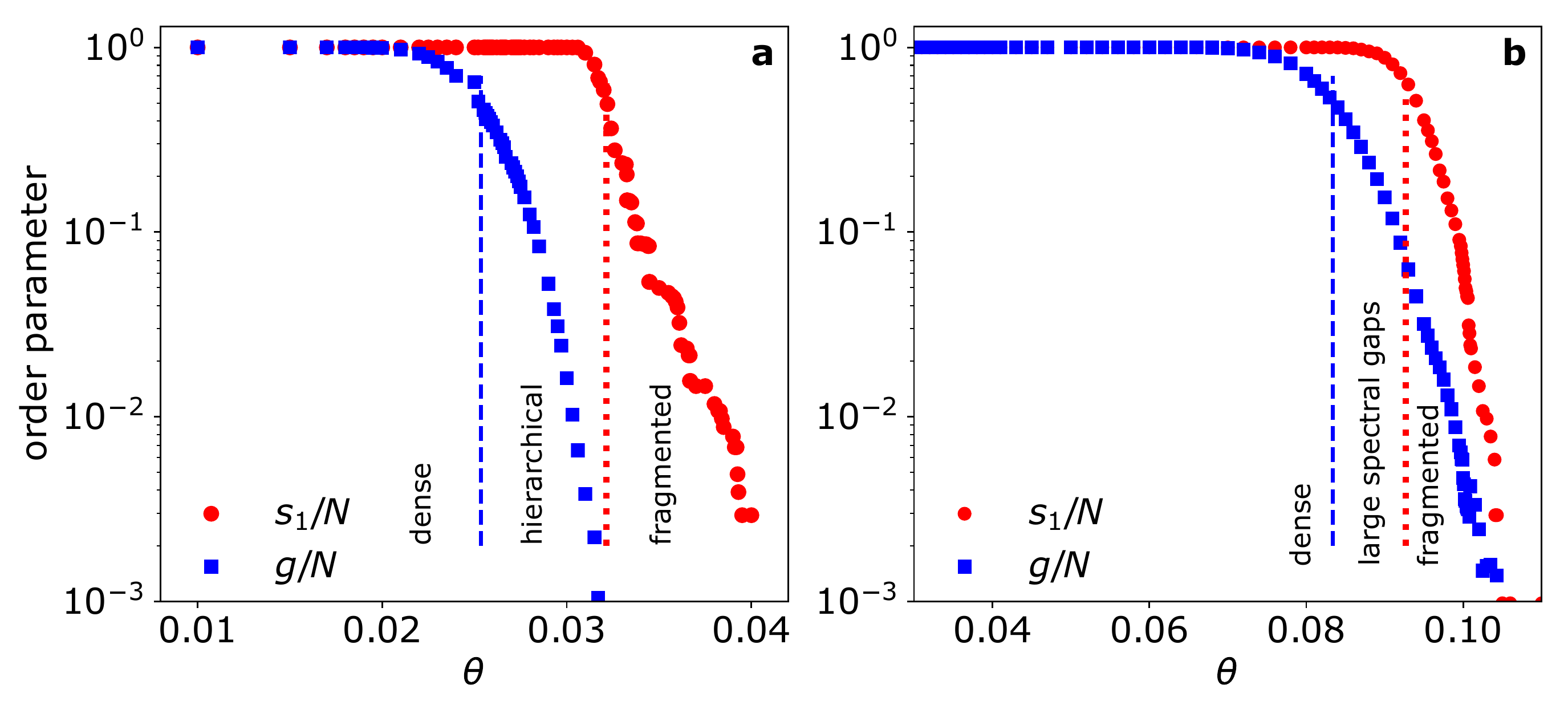}
	\caption{Persistence and breakdown of hierarchical organization. a) Largest connected component size $s_1$ and spectral gap $g$ in the quasi-critical regime ($\kappa\le \kappa_c$), or Griffiths phase, for co-activation matrices $A_{ij}(\theta)$, generated by activity propagation dynamics on a single structural HMN. A range of threshold values $\theta$ exists, where the resulting functional network is connected ($s_1=N$) and has vanishing spectral gaps (by up to three orders of magnitude,  the upper limit given by the system size). More precisely, we construct the range as the interval $[\theta_1,\theta_2]$, such that, given a fixed $f<1$, $g(\theta_1)/N=s_1(\theta_2)/N=f$. Here a we choose $f=0.3$ in order for $\theta_2$ to signal the first significant drop in largest component size: $s_1/N>f$ indicates a network which is still nearly intact; $s_1/N<f$ signals a significant reduction in giant component size, thus pointing to the onset of fragmentation. Different structural networks may produce different $s_1(\theta)$, requiring different choices of $f$.  b) Same as in a) but for simulations run 
in the super-critical regime ($\kappa>\kappa_c$). No significant drop in the spectral gap is recorded while the network is intact, pointing to the loss of hierarchical order. For as long as the network is connected ($s_1=N$), $g$ drops by negligible amounts, and remains of the order of $N$.}
	\label{fig:mainfig}       % Give a unique label
\end{figure}

To produce the functional networks we consider structural HMNs of size $2^{10}$, which already result on dense co-activation matrices of $2^{20}$ thus making the rest of the study computationally intensive. We note that most experimentally acquired connectomes belong in the same size range, allowing for a direct comparison as we will show later on. By running simulations on these synthetic HMNs and acquiring functional data as described in Section \ref{sec:mm},  we observed (see Figure \ref{fig:mainfig}a) that this definition of hierarchical functional networks describes exactly what happens in the quasi-critical regime. Indeed, simulation data for the quasi-critical case show that, upon increasing the threshold $\theta$, the network undergoes a percolation-like transition (red dotted line), as the largest connected component size $s_1$ (red circles) diminishes 
from $N$ (fully-connected network) to $1$ (fully-fragmented network). As we mentionaed in Section \ref{sec:mm}, our main focus is on the spectral gap $g$ and on the way its dependence on $\theta$ differs from that of $s_1$. We find that the spectral gap $g$ (blue squares) too exhibits a transition (blue dashed line), from $N$ (when the functional network is dense and fully connected) to vanishing values. The two transitions however do not coincide and a range of control parameter values $\theta$ emerges, where the network is connected ($s_1=N$), but spectral gaps vanish ($g\to 0$), decreasing by up to three orders of magnitudes. As a guide to the eye, we highlight the $\theta$ interval as the one that begins when $g$ starts decreasing significantly and ends  when $s_1$ starts decreasing by a comparable amount. Thus, within this range, the functional network exhibits a small spectral gap  while still preserving its non-fragmented state; i.e. the network is hierarchical, accordingly to the criterion we proposed.

The strength of this result resides in its dependence on the dynamical regime that we simulate. We just showed that the functional network is hierarchical in the quasi-critical regime, more precisely in the Griffiths phase. But, what happens in the super-critical regime, obtained for activation rates $\kappa>\kappa_c$,  which  is normally associated with abnormal (excessive) neural activity? Figure \ref{fig:mainfig}b shows that in this regime the two phase transitions almost coincide and, more importantly, there is no range of $\theta$ values, where the functional network is connected and the spectral gaps vanish. The functional network is not hierarchical, since whenever it is connected it has large spectral gaps (which only decrease slightly, still maintaining huge values, and not dropping by an amount comparable to $N$).

\begin{figure}
	\centering
	\includegraphics[width=.95\textwidth]{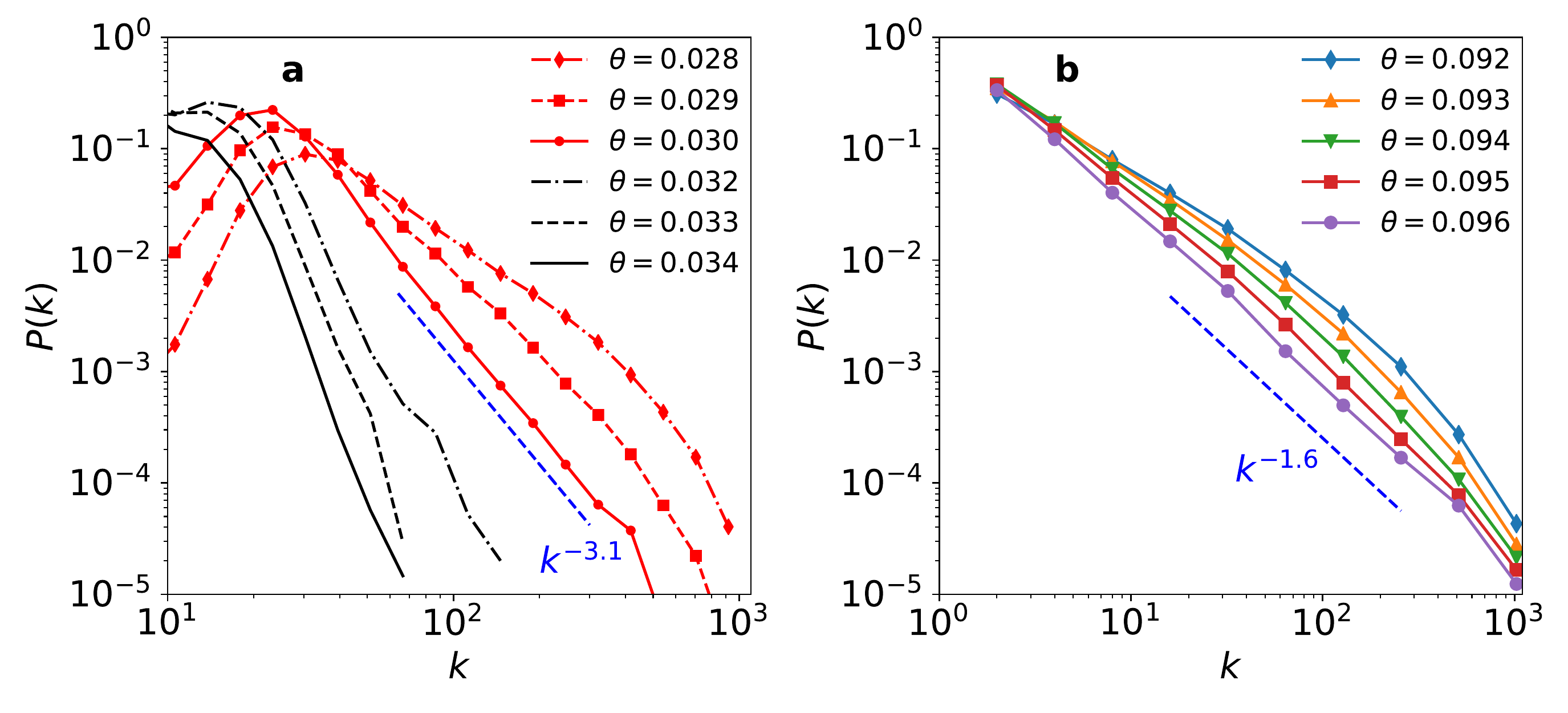}
	\caption{Degree distributions of functional networks. a) Quasi-critical regime. Red curves are obtained for values of $\theta$ producing networks with maximum separation between $s_1$ and $g$. Larger values of $\theta$ (black curves) occur when the network  is close to failure ($s_1\to 0$). In order to improve the statistical sampling, multiple functional networks ($60$) are generated from the same HMN and their degree distributions averaged. The limiting distribution is characterized by an estimated exponent $ -3.1$ b) Super-critical regime. The degree distributions change dramatically, displaying much heavier tails and pointing to a significant rearrangement of the topology with respect to the quasi-critical case. A limiting distribution $P(k)\sim k^{-1.6}$ is encountered for higher values of $\theta$.}
	\label{fig:degreedistributions}       % Give a unique label
\end{figure}

The picture of hierarchical functional networks that develop an emergent, non-hierarchical topology in the super-critical regime is further corroborated by the study of the degree distributions of such networks, as shown in Figure \ref{fig:degreedistributions}. In the quasi-critical phase, and for values of $\theta$ within the range highlighted in Figure \ref{fig:mainfig}a, degree distributions have power-law tails with continuously varying exponents \cite{Clauset2009,Moretti2013}. In the super-critical regime instead, degree distributions display heavier power-law tails with an emergent, limiting exponent close to $3/2$.

\begin{figure}
	\centering
	\includegraphics[width=.5\textwidth]{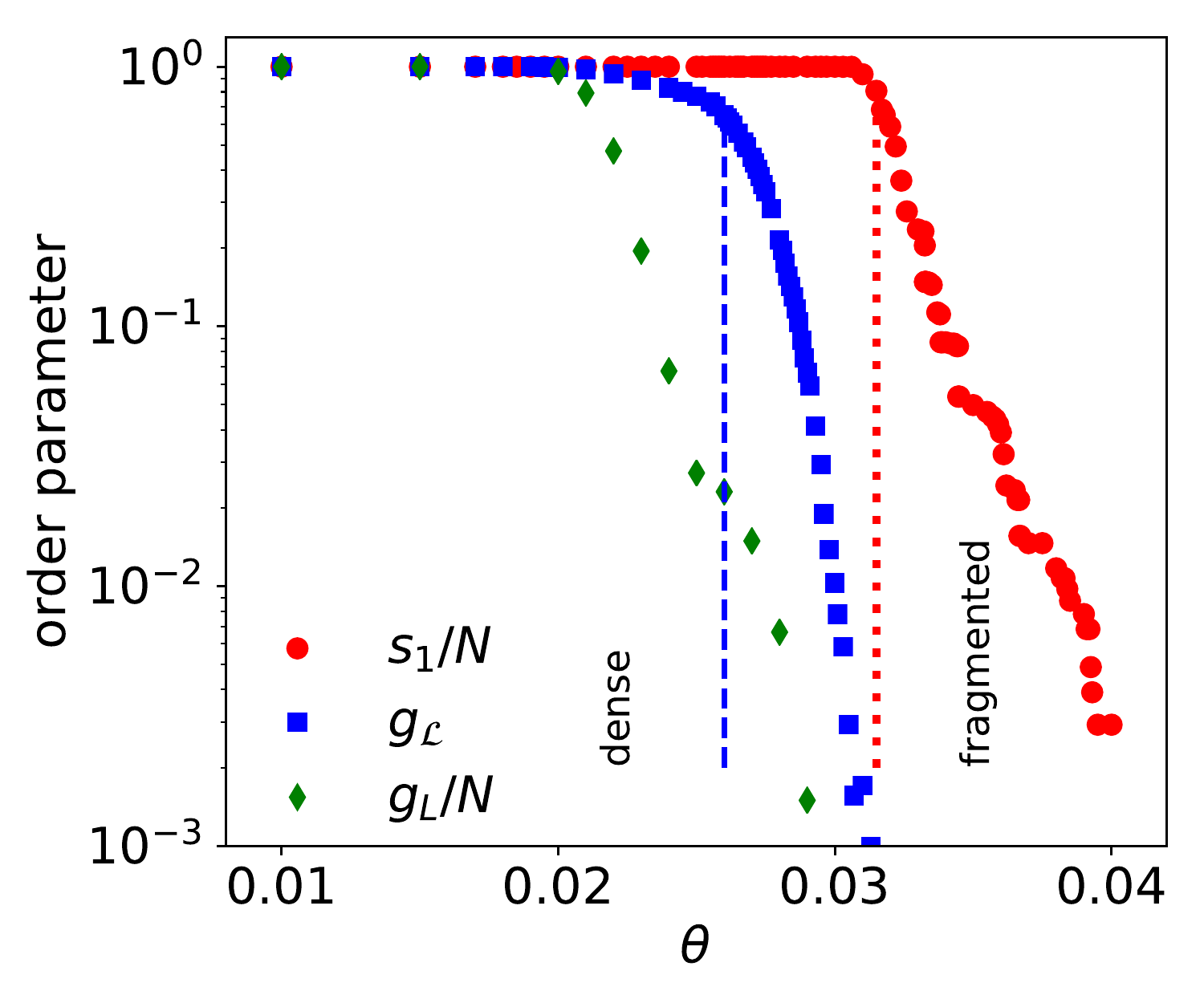}
	\caption{Persistence of hierarchical organization, as from the study of Laplacian matrices. Data is obtained as in Figure \ref{fig:mainfig}a, employing spectral gaps from the graph Laplacian $L$ and the normalized Laplacian $\mathcal{L}$. Data for $s_1$ is the same as in Figure \ref{fig:mainfig}a and is reported here for comparison. We choose $f=0.3$ as in Figure \ref{fig:mainfig}.}
	\label{fig:laplacian}       % Give a unique label
\end{figure}

In order to verify the robustness of our results, we extended the analysis to deal with other forms of connectivity matrices. In particular, our main results so far have been obtained using the definition of spectral gap $g$ applied to the adjacency matrix $A_{ij}(\theta)$. This choice is motivated by the fact that the functional network of adjacency matrix $A_{ij}(\theta)$ is produced by an SIS process, which is linearized by the adjacency matrix $A_{ij}$ of the underlying structural network. Nevertheless, one may wonder if the result above extends to Laplacian matrices \cite{FanChung}, which are of interest in problems of transmission, diffusion and, more importantly in the case of brain dynamics, synchronization \cite{Villegas2014,Serena,Buendia}. Figure \ref{fig:laplacian} shows that our main result carries over to the case of Laplacian matrices, both in the non-normalized variant $L$ (historically, the Kirchhoff Laplacian) and in the normalized one $\mathcal{L}$ (sharing the same spectrum of eigenvalues as the random walk Laplacian).

\begin{figure}
	\centering
	\includegraphics[width=.65\textwidth]{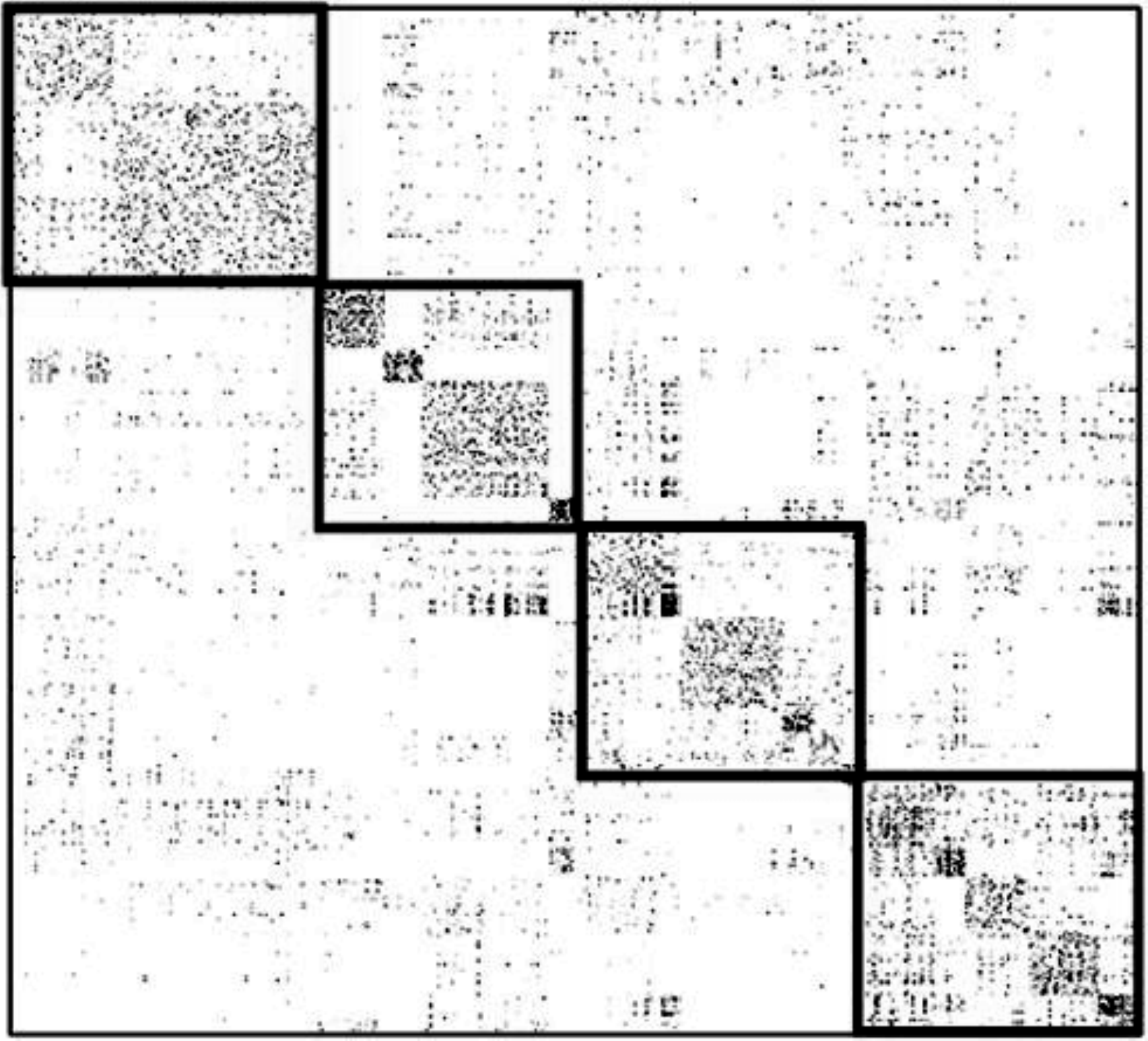}
	\caption{SC adjacency matrix of the human connectome. SC matrices of size $N=2514$ are averaged over $12$ healthy patients. The hierarchical modular organization is highlighted by choosing a node labeling scheme based on agglomerative clustering methods  \cite{Diez2015}. While in our study we employ individual SC matrices, here we show a network average for ease of visualization.}
	\label{fig:brainplot}       % Give a unique label
\end{figure}

\begin{figure}
	\centering
	\includegraphics[width=.95\textwidth]{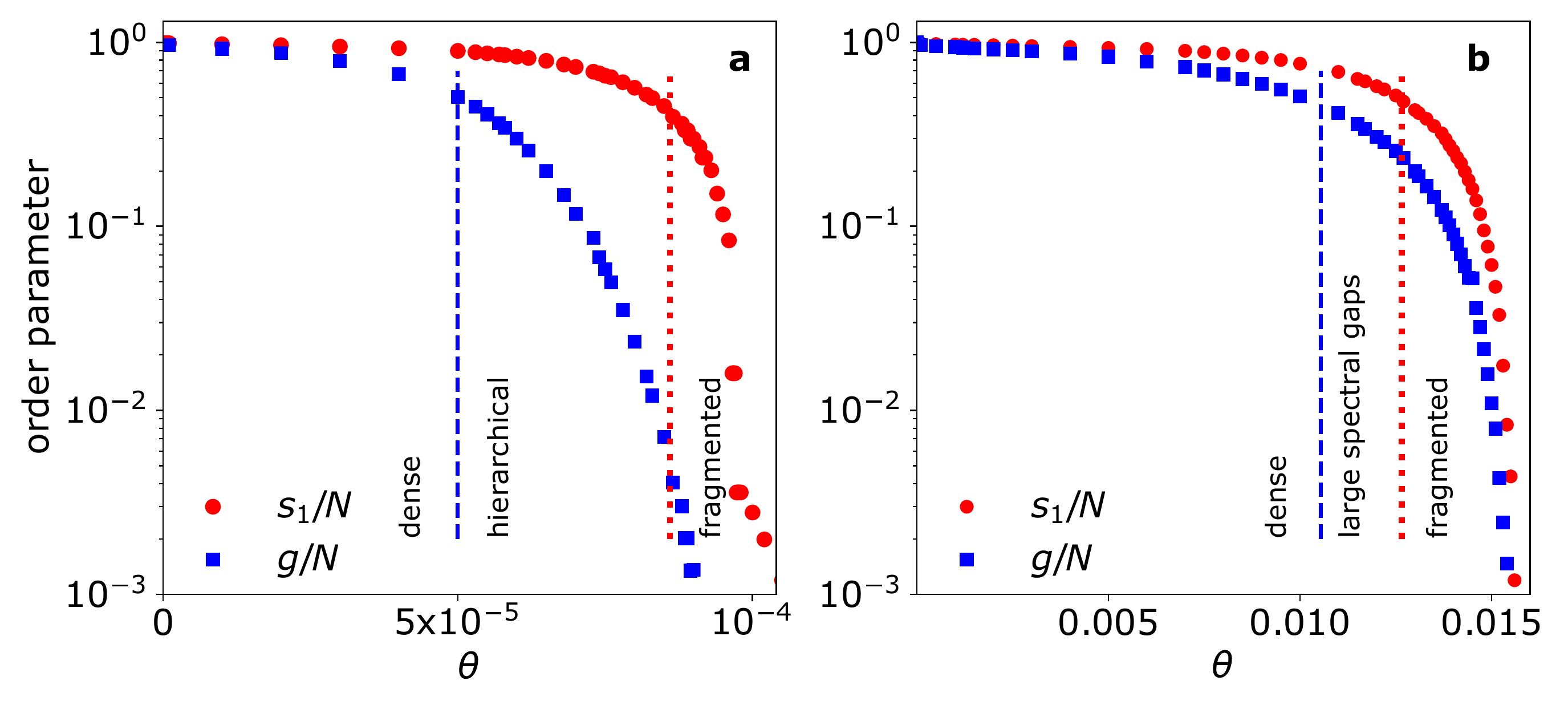}
	\includegraphics[width=.95\textwidth]{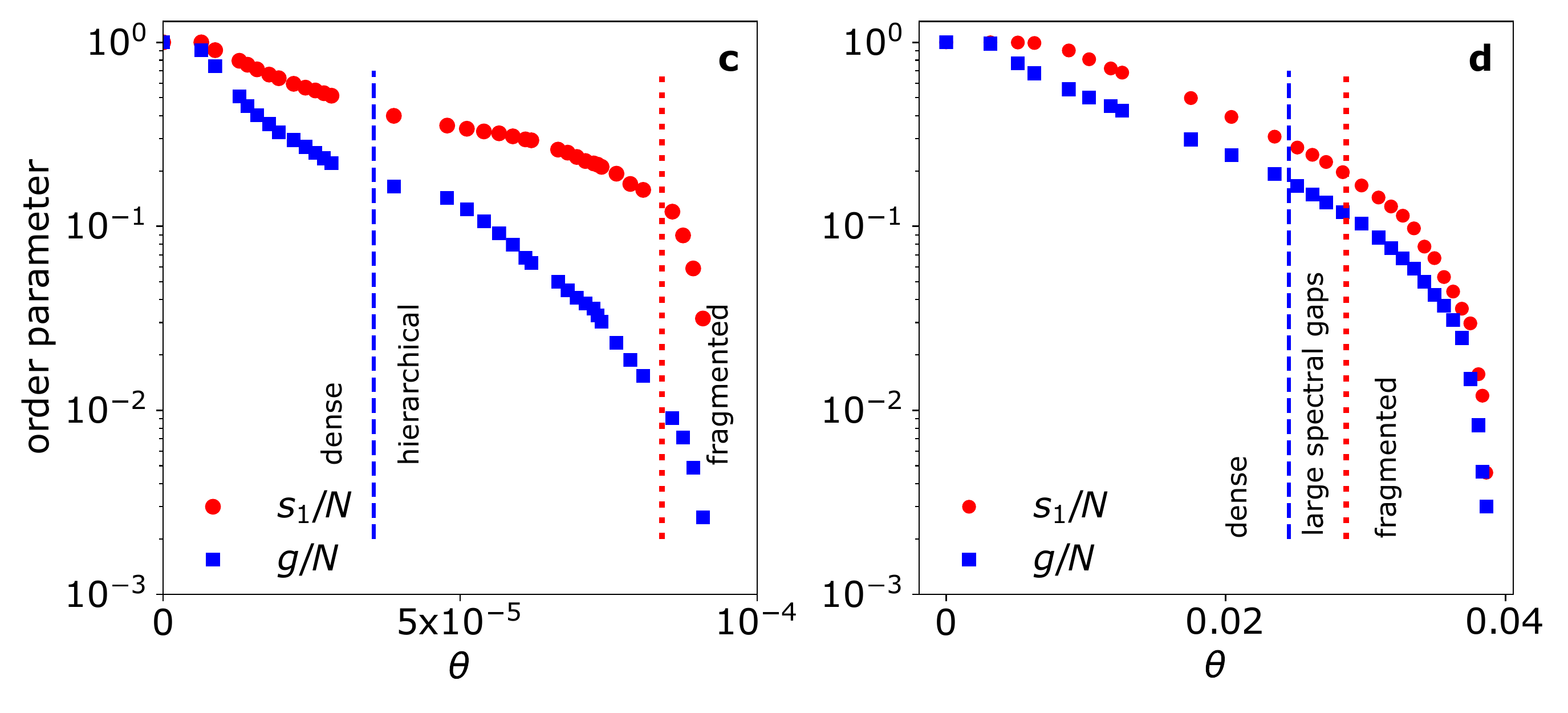}
	\caption{Persistence and breakdown of hierarchical organization of functional networks generated from two real connectomes. (a) and (b):  quasi-critical regime  and super-critical regime for the first subject. (c) and (d):  quasi-critical regime  and super-critical regime for the second subject.  Results are as for the synthetic HMN case shown in Figure \ref{fig:mainfig}. Subject variability introduces changes in the functional form of the curves for $s_1$ and $g$. In particular, in order to highlight the different intervals of $\theta$, we fix $f=0.5$ for the first subject (top) and $f=0.1$ for the second subject, since those are the approximate values of $s_1/N$ for which the networks start fragmenting. Our main conclusions are the same as in Figure \ref{fig:mainfig}. In the quasi-critical regime a range of $\theta$ emerges, where $g$ decreases by over, at least, an order of magnitude while the network is still not fragmented. As for our numerical model, this behavior breaks down in the super-critical case, signaling a pathological state.}
	\label{fig:connectome}       % Give a unique label
\end{figure}

Last but not least, given the relevance of  the above results in the field of computational neuroscience, and in order to prove that they are not artifacts of our specific choice of synthetic HMN networks, we performed the same type of analysis on real human SC connectomes --as described above--  of sizes comparable to the ones considered above ($N=2514$).  The hierarchical modular structure of human connectomes is shown in Figure \ref{fig:brainplot}, were we show the SC adjacency matrix, averaged over 12 healthy patients. We ran our simulations on SC adjacency matrices of individual subjects and performed our spectral analysis of the resulting FC data. Our findings, shown in Figure \ref{fig:connectome} are striking as they show how the same fingerprints of hierarchical functional organization in the quasi-critical case emerge by simulating activity propagation simulations on actual connectomes. Data shown here is produced from the structural network of two healthy subjects (top and bottom row respectively).  

As real connectomes encode  much greater complexity than the HMN models used above, we notice significant changes in the shapes of the $s_1(\theta)$ functions, where an initial slow decrease in $s_1$ anticipates the much faster drop, which we identify with the onset of the network fragmentation. We also notice significant quantitative differences introduced by subject variability. Notwithstanding these differences, the conclusions of our analysis are the same as for the synthetic HMN case: for simulations run in the quasi-critical regime, one can highlight a range of $\theta$ in which $s_1$ is still in its regime of slow decrease, while spectral gaps have dropped by up to two orders of magnitude. A systematic study of a much larger number of subjects is beyond the scope of this paper and is left for future work.

\section{Conclusion and discussion}
The relationship between structure and function in brain networks remains to this day a field of enormous interest. The ultimate goal of devising diagnostic tools that leverage concepts of network theory applied to SC connectivity data begs for a deeper understanding of how dynamic patterns originate from structural motifs, or to which extent they self-organize irrespective of those motifs. What we introduced here is a minimal-model analysis, which however captures some essential aspects of the structure-function relationship. In the quasi-critical regime, activity patterns are strongly localized and result in a FC network which closely mimics the SC hierarchical patterns.  Our criterion to establish the similarity of SC and FC consists in highlighting the existence of a non-trivial threshold interval, in which FC networks exhibit the small-spectral gap property, known to characterize FC networks. Similarity thus consists in the inheritance of a basic spectral property. This is not to say that activity correlations do not play a role in shaping FC patterns: the FC degree distributions do not replicate those of SC, they in fact are power laws with continuously varying exponents. This form of {\it generic} scale invariance has the same structural origin as the one observed in avalanche size measurements in this same regime \cite{Moretti2013,Odor2016}: the hierarchical organization induces rare region effects, which are reflected both in avalanches and in the topology of the functional network.

In the super-critical regime, which represents a pathological state akin to epilepsy, the spectral fingerprint of hierarchical networks is lost and functional connectivity earns a global and much denser network topology. While it was shown in the past that networks with localized structures host localized activity patterns also in the super-critical phase \cite{Goltsev2012} ---something we were able to confirm in HMN models of brain activity \cite{Safari2017}--- the change in spectral properties and degree distributions reported here points to a significant reorganization of dynamics and correlations, in which modules mutually reactivate %(or mutually reinfect, as in similar models of epidemics  \cite{Boguna2013,Wei2017}.

Our methodology addresses the analysis of activity correlation data without fixing an arbitrary threshold. Thresholds are treated as a control parameter in a phase diagram, which is analyzed as a whole, to show how, in the quasi-critical regime, threshold values which correspond to non-trivial functional network topologies (fully connected, fragmented) also lead to the non trivial property of small spectral gaps, which allows us to conclude that the functional networks effectively inherit the hierarchical organization of the structural ones.  
 
We believe that our work serves as a proof-of-concept study, for more complex applications in which multiple spectral indicators are considered, and in which FC is inferred from real time series, rather than from numerical simulations. For instance, by altering the natural dynamics by drugs able to block excitation and/or inhibition, one could analyze how and when those alterations are reflected in the hierarchical structure of FC networks and in its possible breakdown.

The simple methodology we adopt here, based on co-activations, has clear advantages while generating functional networks under minimal assumptions, and is motivated by the simplified nature of the dynamic model in use here. At the same time, our study can also be conducted on FC data extracted as Pearson correlation, allowing for instance the separation of correlation and anti-correlation effects. Given the fact that both structural and  functional data is available to us for a large number of subjects, as detailed in Section \ref{sec:mm}, a large-scale application of our methodology is planned an the next step of our investigation.  In passing, we note that since the present results can be obtained with data acquired in a short time window, our approach also lends itself naturally to the study of experimental data sets obtained in the form of dynamical FC.

Our work of course relies on essential simplifications of the original neuroscience problem. Namely our simulated dynamics is a simplification of neural activation processes. Nevertheless, we believe that the results reported here provide useful insights regarding the extent to which structure and function are connected in brain networks, and how pathological states can be accompanied by radical shifts in functional connectivity. We believe that the type  of spectral analyses presented here to study persistent and emergent topologies in FC data sheds light on the relationship between optimal architectures and tuneable performance in the broader context of neurocomputing.

%\bibliography{mybibfile}

\section*{Appendix}

Data available to us are from $30$ healthy subjects ($14$ males, $16$ females) with age between $22$ and $35$. Data were provided through the Human Connectome Project, WU-Minn Consortium (Principal Investigators: David Van Essen and Kamil Ugurbil; 1U54MH091657) funded by the 16 NIH Institutes and Centers that support the NIH Blueprint for Neuroscience Research; and by the McDonnell Center for Systems Neuroscience at Washington University.  To build the connectivity matrices (for  further details see   \cite{serafim2019}),  we processed    the  same-subject structure-function triple acquisitions of magnetic resonance imaging (MRI) -- see Appendix for details on acquisition parameters -- consisting of:   1. High-resolution anatomical MRI (used for the mask segmentation of     gray matter, white matter and cerebrospinal fluid, and for  the transformation to common-space of the functional and the diffusion data), 2. Functional MRI at rest (used for extracting region time series of the blood-oxygen-dependent signal, after removal of movement artifacts and physiological noise, but not the global signal regression), and 3. Diffusion MRI (used for building SC matrices after fitting a diffusion tensor to each voxel,   running a deterministic tractography algorithm  using the UCL Camino Diffusion MRI Toolkit \cite{Camino}, and counting  the number of streamlines connecting all pairs within the $N=2514$ regions,  each one containing on average 66 voxels.

\subsection*{Imaging acquisition parameters}

Same subject structure-function triple acquisitions were performed using a 3T Siemens Connectome Skyra with a 100 mT/m and 32-channel receive coils. The acquisition consisted of\footnote{Here, we only report the parameters which are neeeded for the image preprocessing, but a complete information about the acquisitions can be found at \url{http://protocols.humanconnectome.org/HCP/3T/imaging-protocols.html}}:

\textit{High-resolution anatomical MRI}  was acquired using a T1-weighted 3D MPRAGE sequence with the following parameters:  
TR=2400ms; TE=2.14ms;  Flip angle=8deg;  FOV=224$\times$224 mm$^2$;  Voxel size=0.7mm isotropic; Acquisition time = 7 min and 40 sec.

\textit{Functional data at rest} were acquired to obtain the   blood-oxygenation-level-dependent (BOLD)   signals with  a gradient-echo EPI sequence with the following parameters: TR = 720ms, TE = 33.1ms; Flip angle = 52 deg; FOV = 208 $\times$ 180 mm$^2$; Matrix = 104 $\times$ 90; 72 slices per volume, a total number of 1200 volumes;   Voxel size =  2mm isotropic; Acquisition time =  14 min and 33 sec.

\textit{Diffusion data} were acquired with a Spin-echo EPI sequence with the following parameters: TR = 5520ms; TE = 89.5ms; Flip angle = 78 deg; FOV = 210 $\times$ 180 mm$^2$; Matrix = 168 $\times$ 144; 111 slices per volume;  Voxel size =    1.25mm isotropic;   90 diffusion weighting directions plus 6 b=0 acquisitions; Three    shells of b=1000, 2000, and 3000 s/mm2; Acquisition time: 9 min 50 sec.

\section*{Abbreviations}
FC: functional connectivity
HMN: hierarchical modular network
MRI: magnetic resonance imaging
SC: structural connectivity
SIS: susceptible-infected-susceptible

\section*{Competing interests}
The authors declare that they have no competing interests.

\section*{Ethics approval and consent to participate}
The sample included 30 subjects from the MGH-USC Human Connectome Project. The research was performed in compliance with the Code of Ethics of the World Medical Association (Declaration of Helsinki). All subjects provided written informed consent, approved by the ethics committee in accordance with guidelines of HCP WU-Minn.

\section*{Funding}
We acknowledge the Spanish Ministry and Agencia Estatal de investigaci{\'o}n (AEI) through grant FIS2017-84256-P (European Regional Development Fund), as well as the Consejería de Conocimiento, Investigación  Universidad, Junta de Andalucía and European Regional Development Fund, Ref. A-FQM-175-UGR18 and SOMM17/6105/UGR for financial support.
AS and PM acknowledge financial support from the Deutsche Forschungsgemeinschaft, under grants \mbox{MO 3049/1-1} and \mbox{MO 3049/3-1}. 

\section*{Availability of data and materials}
Numerical simulation data are available on request from the corresponding author.

\section*{Author contributions}

AS: Performed the simulations, Analyzed the data, Wrote the manuscript; PM: Designed the study, Analyzed the data, Wrote the manuscript; ID: Data curation, Preprocessed the images; JMC: Data curation, Wrote the manuscript.  MAM: Designed the study, Analyzed the data,  Wrote the manuscript

\section*{Acknowledgments}

\end{document}